\documentclass[eqsecnum,aps]{revtex4}
\usepackage{graphicx,color}
\newcommand{\be}{\begin{equation}} 
\newcommand{\ee}{\end{equation}} 
\newcommand{\bea}{\begin{eqnarray}} 
\newcommand{\eea}{\end{eqnarray}}
\newcommand{\G}{\Gamma_4}
\newcommand{\D}{\Delta_4}
\newcommand{\K}{K_{d+1}}
\newcommand{\lb}{\left (}
\newcommand{\rb}{\right )}
\newcommand{\mb}{\mathbf}
\newcommand{\mA}{\mathcal{A}}
\newcommand{\figyz}
{\begin{figure}[htbp]
        \centering

        \includegraphics[angle=0,width=6.5cm]{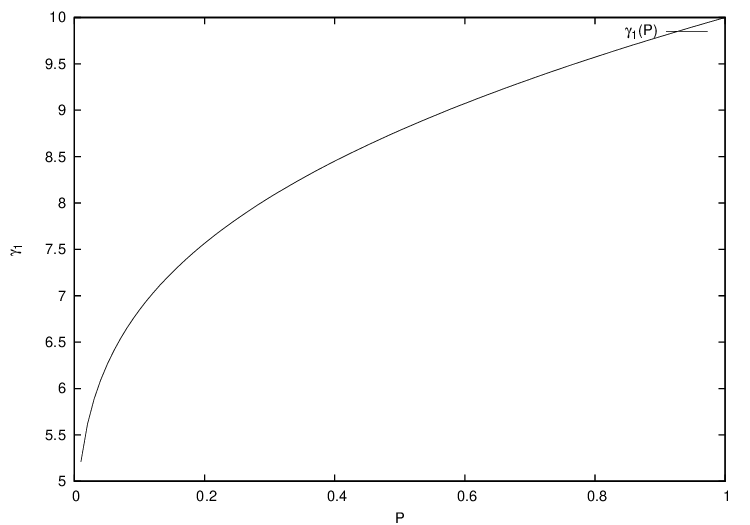}
           \caption{Asymptotic evolutions of $\gamma_1$  with $P$ at $T=0$. 
                    Parameter values are $3\G^0 =10$ and $a=0.1$.}
\label{zero}
	\end{figure}
} 
\newcommand{\figT}
{\begin{figure}[htbp]
        \centering
         \includegraphics[angle=0,width=6.5cm]{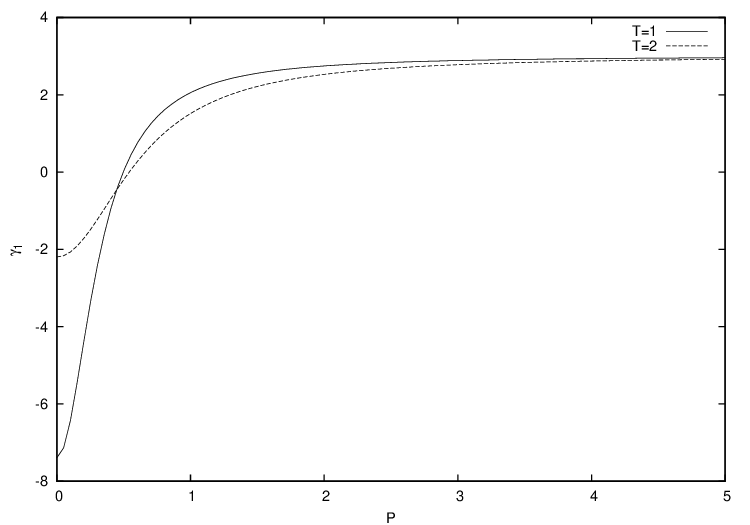}
           \caption{Asymptotic evolution of $\gamma_1$  with $P$ at  $T\ne0$.
                    Two curves are drawn at two different temperatures with 
                    $\G^0 =10/3,\, a=0.1$ and $K_d=0.1$. Upper one is at $T=1$ and the lower one is 
                    at $T=2$ in some arbitrary scale.}
\label{finite}
	\end{figure}
} 
\newcommand{\figvrtx}
{\begin{figure}[htbp]
        \centering
        \includegraphics[angle=0,width=7cm]{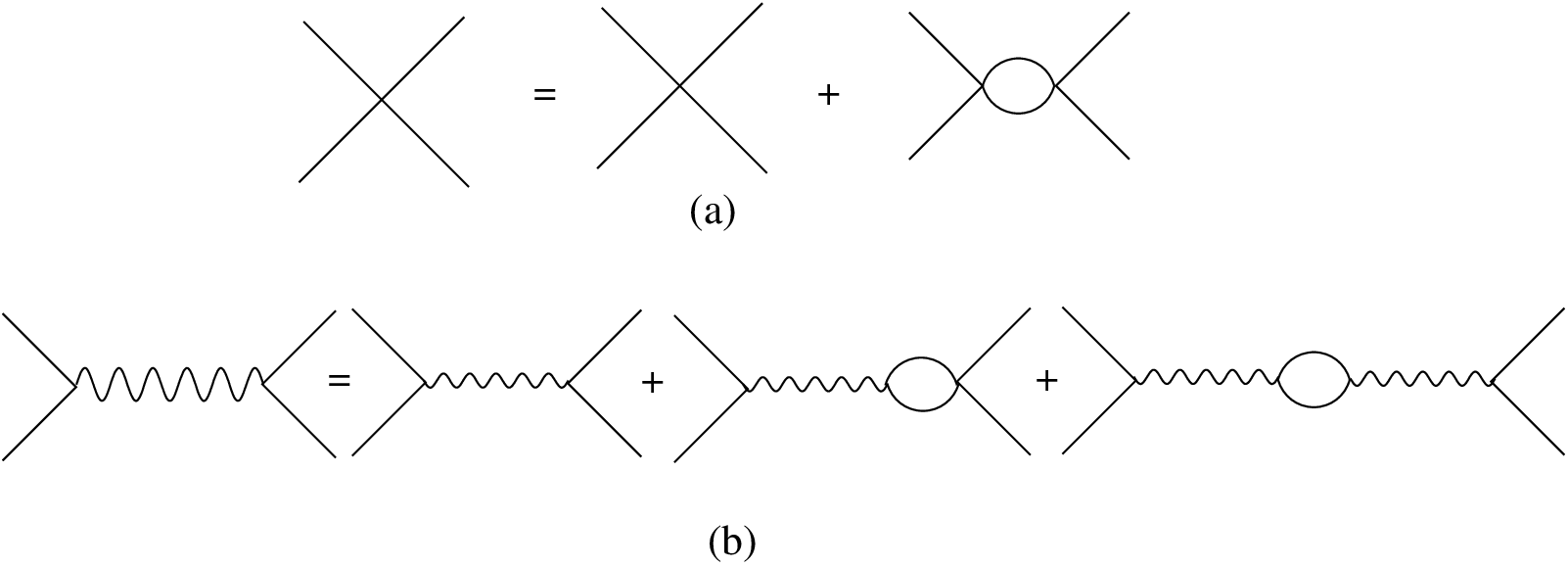}
		\caption{Parquet diagrams for the fluctuation corrections to the short interaction vertex $u$. 
                 It takes care of the loop corrections of the above type to all orders. Here each line corresponds 
                 to the propagator given by the equation[\ref{prop}]. }
\label{f-parquet}
	\end{figure}
}
\newcommand{\figGap}
{\begin{figure}[htbp]
        \centering
        \includegraphics[angle=0,width=7cm]{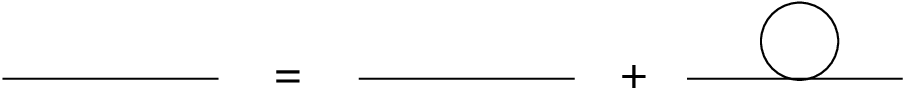}\hspace{0.75cm} \
		\caption{Diagrammatic representation of  gap renormalization up-to one loop. }
\label{gap}
	\end{figure}
}
\begin{document}

\title{Effects of Strain coupling and Marginal dimensionality in the nature of phase transition in Quantum paraelectrics}
 
\author{Nabyendu Das$^{1,2}$ }


\affiliation{$^1$Institute of Physics, Bhubaneswar 751005, India\\ 
$^2$Saha Institute of Nuclear Physics, Kolkata 700062, India.}

\date{\today} 

\begin{abstract}
       Here a recently observed weak first order transition in doped SrTiO$_3$\cite{Taniguchi} is argued to be a consequence 
       of the coupling between  strain and  order parameter fluctuations. Starting with 
       a semi-microscopic action, and using renormalization group equations for vertices, 
       we write the free energy of such a system. This fluctuation renormalized free energy 
       is then used to discuss the possibility of first 
       order transition at zero temperature as well as at finite temperature. An asymptotic 
       analysis predicts small but a finite discontinuity in the order parameter near a mean field quantum 
       critical point at zero temperature. In case of finite temperature transition, near 
       quantum critical point such a possibility is found to be extremely weak. Results are in accord with some experimental findings on 
       quantum paraelectrics such as SrTiO$_3$ and KTaO$_3$.
\end{abstract}
\maketitle
\section{Introduction}
The static dielectric susceptibility in the materials like SrTiO$_3$ and KTaO$_3$
 does not show any divergence down to 0K, rather saturates at a 
very high value $(\mathcal{O}(10^4))$. Ferroelectric transitions in materials 
which are similar in structure,  for example in BaTiO$_3$, are of displacive type 
and are well described by softening of a zone center transverse optical mode at some 
finite temperature. This scenario, that is, the absence of transition and the saturation at 
high dielectric susceptibility, has been attributed to the smallness of the zero temperature 
gap in the corresponding optical branch. Smallness of the gap makes quantum fluctuations 
relevant in these materials. These materials are called quantum paraelectrics
\cite{Muller}. There is a revival of interest, particularly regarding the nature of phase 
transition at and near quantum critical point in these materials \cite{Millis,we,Coleman}. 
The effects of pressure\cite{expt-pressure} and impurities on the dielectric susceptibilities 
have been well studied experimentally. One recent experimental report\cite{Taniguchi}, based on
Raman spectroscopic study,
indicates that these materials show phase separation near a zero temperature instability 
and disappears as one moves toward finite temperature. Moreover the intensity of the signals
indicating ferroelectric order in a paraelectric background is very weak. Such a finding is a clear
indication of a first order transition where discontinuity in order parameter is small and hence its 
fluctuation is important. 
Apart from that a general 
consensus arising form other experiments is that the application of hydrostatic pressure  
moves them away from criticality and the possibility of phase transition is suppressed. 
One needs to apply, what is termed as, \emph{a negative pressure} to induce phase transitions in 
these materials. One way to simulate negative pressure is to put non-polar impurities 
which creates local pressure deficiencies. In this context, experimentally
\cite{expt-impurity} one finds $T_c \sim (n-n_c)^{\frac{1}{2}}$ 
(where n is the average impurity concentration and $n_c$ is the critical value, typically 
$33\%$) which matches well with the theoretical estimated\cite{we} transition temperature 
for pressure induced transition. In this case, the mean field $T_c \sim (p+p_c)^{\frac{1}{2}}$ 
(where $p$ is hydrostatic pressure and $p_c$ is the critical value). The exponent $\frac{1}{2}$ 
is obtained when thermal fluctuations are treated at the Gaussian level. The similarity between 
effects of pressure and the impurity in the transition temperature can be attributed to 
the high density of impurity concentrations in these cases. Here the disorder effects seem to be 
small and a non-polar impurity essentially induces an internal pressure. This motivates us to develop
a description, suitable for the properties of the pressure induced phase transition, which can be used to 
understand the occurrence of the phase separation mentioned above in the ferroelectric 
transition near a ferroelectric quantum critical point. In this work, we consider the displacive type 
ferroelectrics in its quantum limit. In these materials the atomic dipoles are believed to 
interact with each other via a {\it long range dipolar interaction} and their collective excitations give
rise to various optical modes. The long range nature of the dipolar interaction affects the longitudinal
optical modes and makes them rigid and they do not contribute to the low temperature dynamics.
Hence to look for a weak first order transition scenario (correlation length is still large), we start with
 short range interaction without dipolar anisotropy and try to explore the results of order 
parameter fluctuations with an effective long range interactions 
among them, mediated by strain fluctuations. Without altering the basic nature of the transition mechanism in these systems, we start 
with an one phonon band  model and look for the fluctuation effects in four point vertices. To retain 
the leading fluctuation effects in the vertex function along with their dependencies on the non-zero polarization, 
we calculate the free energy using well known RG equations. This fluctuation renormalized free energy 
indicates possibility of a first order transition at zero temperature and it also gives finite temperature 
corrections. 
\section{Summary of the mean field analysis}
We assume that a fluctuating strain field $\epsilon_{ij}(\tau)\sim (\nabla_i u_j (\tau)+ \nabla_j u_i(\tau))$ couples to a bi-linear form of the optical mode fluctuations as $g\epsilon_{ij}(\tau)\phi_i(\tau)\phi_j(\tau)$. Here $u_i(\tau)$ represents the displacement due to acoustic mode fluctuations at $i$-th site in real space and $g$ is the opto-elastic coupling. We consider the Gaussian fluctuations of the strain fields and consider the system to be at a constant pressure. Integrating out the strain fluctuations completely, we get an effective long range interactions among the optical mode fluctuations of the form $v\phi^2_i\phi^2_j$ in real space.   Here $v\propto g^2$ and depends on various elastic constants depending on which it can be either negative or positive. A naive quantum generalization of such interaction would lead to a term like $v\phi^2_i(\tau)\phi^2_j(\tau)$, where $\tau$ is time. Such a term indicates field variables at two different position interact at same point with same interaction 
strength. This clearly violates causality and we need to introduce non-locality in time in such interaction. Thus we the resulting interaction to be $v\phi^2_i(\tau_i)\phi^2_j(\tau_j)$ which is found to consistent with the quantum-classical mapping of our strain coupled system. Since we are considering a weak first order transition a-priori, we assume $v$ to be negative and leave the detailed discussions on its dependence on various elastic constants.
 In a Fourier space our effective action describes only polarization fluctuations transverse to the momentum vector with strain induced long range interactions among them and takes the following form,
\bea
\mA &=& \frac{1}{2\beta}\sum_{ q} (\omega_n^2 + r+cq^2 )\phi_q \phi_{-q} 
+ \frac{u}{4!\beta}\sum_{ q_1, q_2, q_3} \phi_{q_1}\phi_{q_2}\phi_{q_3}\phi_{-q_1-q_2-q_3}
\nonumber \\ 
&& + \frac{vL^{-d}}{4!\beta}\sum_{ q_1, q_2 } \phi_{q_1}\phi_{-q_1}\phi_{q_2}\phi_{-q_2}.
\label{H_0}
\eea
Here $\phi_{q_i}=\phi(\mb{q}_i, \omega_{n_i})$ describes the Fourier transform of  local transverse polarization, $\mb{q}$ is the field momentum, $\omega_n={2\pi n}/{\beta}$ is 
the Matsubara frequency for Bosonic excitations,  $r$ and $u$ are the coupling constants for quadratic and the anisotropic short
range quartic interactions respectively. The parameter $v$ is the coupling constant for isotropic long range part of the quartic coupling 
induced by strain and $L^{d}$ is the system volume in $d$-spatial dimension. Hydrostatic 
pressure, as well as the non-polar impurity, couples to the optical mode via strain. It shifts the bare quadratic 
and quartic coupling $r_0$ by $r=r_0(1+p/p_0)$, where $p$ is the homogeneous pressure and $p_0$ is a constant. Strain fluctuations induce a long range attractive interaction between the dipoles and is denoted by the effective quartic coupling $v$.  We focus on a weak first order transition near a quantum critical point where $\phi$ acquires a non-zero value. Thus in a mean field approximation near such a transition point $\phi$ can be decomposed into two parts, $P$ the static mean field part and $\psi(\mb{q}, \omega)$, the fluctuating part as follows,
\be 
\phi_q = P\delta(\mb{q},\omega) + \psi(\mb{q}, \omega)
\ee  
It is assumed that $<\psi(q, \omega)>=0$. In this approximation our starting action (\ref{H_0}) can be 
rewritten as
\bea
\mA &=& \frac{r}{2} P^2 + \frac{(u+v)}{4!}P^4+\frac{1}{2\beta}\sum_{q} (\omega_n^2 + r+cq^2 + (u/2+v/6)P^2)\psi_{q}\psi_{-q} \nonumber \\
&&+P\frac{u}{3!\beta}\sum_{q_1, q_2}\psi_{q_1}\psi_{q_2}\psi_{-q_1-q_2}+\frac{u}{4!}\sum_{ q_1, q_2, q_3}\psi_{q_1}\psi_{q_2}\psi_{q_3}\psi_{-q_1-q_2-q_3}
\nonumber \\
&&+\frac{vL^{-d}}{4!\beta}\sum_{ q_1, q_2 }\psi_{q_1}\psi_{-q_1}\psi_{q_2}\psi_{-q_2} .
\label{H-fluc}
\eea
Here we use the notation $\psi_{q_i}=\psi(\mb{q}_i, \omega_{n_i})$. It is to be noted that the term $P^2 \psi_{q}\psi_{-q}$ has a coefficient $3u+v$ which can remain positive even when $u+v<0$ and it has important consequences which will be discussed later. Technically the long range part of the action with vertex $v$ contributes to such term two possible ways whereas the short range part with vertex $u$ contributes in six $(C^4_2)$ possible ways and the difference lies in their range of interactions. With this action we can study the thermodynamics of the system by constructing a free energy which is defined as logarithm of a functional integral over $\mA(\psi, P)$, i.e.
\be
\mathcal{F} = -\frac{1}{\beta}\log\lb\int \mathcal{D}\psi e^{-\mA(\psi, P)}\rb.
\ee
The value of $P$ is to be determined by minimizing the free energy $\mathcal{F}$. Stability of a thermodynamic system 
requires the free energy to be positive. In a Landau theory, 
which neglects fluctuations completely, stability criteria requires the coefficient of the quartic term, i.e. $(u+v)$  to be positive. In that case, for $r>0$, the free energy will be minimized for $P=0$ resulting in a second order transition. On the other hand for $(u+v)<0$ stability criteria in a mean field theory requires a higher order term with positive coefficient
which results a first order transition with a non-zero $ P\sim |u+v|$. We consider a limiting situation where $|u+v|\approx0$ which corresponds to a weak first order transition. In this regime a proper account of the fluctuation corrections should be taken and it will be shown that fluctuation corrections alone can stabilize the system without invoking a higher order term in the starting action. We will discuss the effects of fluctuations in four point vertices near a weak first order transition in the next section. 

\section{Fluctuation corrections to the free energy at zero temperature} 
In the previous section we discussed importance of the coefficients of quartic term to determine the nature of phase transition. In a field theory description these coefficients are called vertex functions. Under certain circumstances they
can get heavily renormalized by order parameter fluctuations and an weak first order transition is such an event. In this case there is a competition between order parameter fluctuations and a non-zero value of  order parameter to stabilize a thermodynamic system. To quantify  the effects of the competition between the order parameter fluctuations and a non-zero value of  order parameter we calculate the fluctuation re-normalized four point vertex functions. Then a fluctuation renormalized free energy is constructed using them. We use renormalization group equations for four point vertices obtained in the lowest order perturbation theory. Such equations were derived earlier by
Gadeker and Ramakrishnan\cite{Gadeker,GadeThs}  in a \emph{parquet approximation}. It is assumed that near a weak first order transition, a system acquires a small but non zero polarization $P$, the polarization fluctuation near such a phase transition becomes gapped, with the gap being proportional to  $P^2$. Thus near such a phase transition the free optical phonon propagator which is the inverse of the coefficient of the quadratic term of the fluctuating part in the mean field action (eqn. (\ref{H-fluc})), is given by,
\be
G^{-1}(q,\omega_n)= r+cq^2+\omega^2_n + (u/2+v/6)P^2 .
\label{prop}
\ee
\figvrtx
In a paraelectric phase with $u+v>0$, $P=0$. A bare theory predicts that at $T=0$, the 
static susceptibility $\chi(0,0)\sim G(0,0)\sim 1/r$. Thus $r=0$ is a point of instability in the paraelectric phase, and when  there is no discontinuity in the order parameter ($u+v>0$) at that point, it can be identified as a quantum critical point. In the vicinity of the critical point, correlation length  becomes large, leading to dominance of the order parameter fluctuations. Moreover when $|u+v|\approx 0$, fluctuation corrections to the four point vertices become important. We will try to discuss the effects of fluctuations in the vicinity of the limit $|u+v|\rightarrow 0$, in developing  spontaneous non-zero value of the order parameter near the transition 
point. Since in the case of a week first order transition, correlation length grows significantly, the bare vertices get strongly re-normalized.  
Now using the bare propagator for the order parameter
fluctuation (eqn. (\ref{prop})) we find 
that the leading order contribution from the second diagram of the figure \ref{f-parquet}(a) to the renormalization vertex function with a momentum cutoff $\Lambda$  is given as, 
\be
\delta u =  -u^2 \sum_n \int_0^\Lambda d^dq G^2(q, \omega_n).
\ee
 At zero temperature the frequency summation becomes an integral and in this case, the combination of the frequency sum and the $d$-dimensional 
integral can be replaced by a $d+1$ dimensional integral. Hence in three dimension, 
\bea
\delta u &\sim&   -u^2  \int_0^\Lambda d^{3+1}q G^2(q)~{\rm which}\nonumber\\
&\sim&  -u^2 K_4\log\frac{\Lambda}{(r+ (u/2+v/6)P^2)^{\frac{1}{2}}}.
\eea
 Here $K_4$ is a constant which is related to the surface area of a $4$-dimensional sphere of unit radius. Surface area of a d-dimensional hyper-sphere is given by,
\bea
K_d = \lb 2^{d-1}\pi^{\frac{d}{2}}\Gamma(d/2)\rb^{-1}.
\eea
$ K_d = \frac{1}{8\pi^2}$ and $\frac{1}{2\pi^2}$ in d=4 and d=3 respectively.
The correction to four point vertex $\delta u$ has a logarithmic divergence as $r\rightarrow 0$ and $P\rightarrow 0 $. We define the diverging 
logarithmic part as a new cut-off variable
\be 
x=\log\frac{\Lambda}{(r+ (u/2+v/6)P^2)^{\frac{1}{2}}}.
\label{parquet}
\ee
 We define the re-normalized proper four point vertices  $\G$ and $\D$, with their bare values given as,
\be
\G^0 = u  ~{\rm and}~ \D^0 = v.
\ee
Considering the lowest order corrections, we get the following renormalization group equations in terms 
of the cut-off variable $x$,
\bea
\frac{d\G}{dx} &=& -\frac{3}{2} \K \G^2 (x),
\label{self-eqn1}\\
\frac{d\D}{dx} &=& -\K \G (x)\D (x)- \frac{1}{6} \K\D^2(x).
\label{self-eqn2}
\eea
The above equations can also be obtained in a parquet re-summation scheme by summing leading order diagrams up-to infinite order as shown in figure \ref{f-parquet}. However the solutions of these equations can be written as,
\bea
\G &=& \frac{\G^0}{1+ \frac{3}{2}\K \G^0 x},\nonumber\\
\D &=& \frac{3\G\D^0}{\D^0 
+(3\G^0-\D^0)(1+ \frac{3}{2}\K \G^0 x)^{-1/3}}.
\label{gamma-delta}
\eea
In this derivation the contributions from the third order term, i.e. from $P\psi\psi\psi$ is neglected. In a perturbative theory, this term contribute nothing at the first order. It contributions to the higher order. But those corrections are less divergent compared to the contributions coming from quartic terms. It is to be noted that the cut-off variable $x$ contains $\G$ and $\D$. Thus the set of equations (\ref{gamma-delta}) defines coupled equations for $\G$ and $\D$. They need to be solved self-consistently 
to find out their dependencies on $r$ and $P$. From $\G$ and $\D$ thus obtained, we can calculate the fluctuation re-normalized free energy using the relation,
\be 
\frac{\partial^4 F}{\partial P^4}=\G +\D.
\label{identity}
\ee  
We need to integrate (with proper boundary conditions) the above equation four times with
respect to $P$ to get an expression for the free energy. 
Integrating the equation (\ref{identity}) once, we get
\bea
\frac{\partial^3 F}{\partial P^3} &=& \int_0^P (\G +\D) dP'\nonumber\\
&=&  P(\G +\D)+\int_0^P P' \frac{\partial}{\partial P'}(\G +\D) dP' + c(r).
\eea 
Here $c(r)$ is  a constant independent of $P$ and can be equated to zero using the symmetry constraint ${\partial^3 F}/{\partial P^3}|_{0}=0$. The second term in the right hand side of the above equation takes care of the $P$ dependence of $\G$ and $\D$. 
To make our calculations simpler, we will neglect that term at this stage. 
Before doing so, we make an estimate of the corresponding error. The integral reads as,
\bea
\int_0^P\frac{2\K(3\G+\D)^3 P'^2 dP'}{6r+[(3\G+\D)-\frac{\K}{6}(3\G+\D)^2]P'^2}.
\label{corr}
\eea
Here $\K =\frac{1}{8\pi^2}$ in $d =3$. In order to make our lowest order perturbation theory valid, we choose $(3\G +\D) \sim\mathcal{O}(10)$. Thus $\K (3\G +\D) \sim \mathcal{O}(10^{-1})$.
The contribution from the $P$-dependence of $\G$ and $\D$ to the integral(\ref{corr}) becomes,
\bea
&&\frac{\K}{3} (3\G +\D)^3\int_0^P \frac{P'^2 dP'}{6r+(3\G +\D)P'^2}\nonumber\\
&\approx&\frac{\K}{3} (3\G +\D)^2P \approx 10^{-2}(3\G +\D)P.
\eea
We see that, the $P$-dependence of $\G$ and $\D$ contributes two order of magnitude less compared to the other terms in the free energy. 
Thus we neglect the $P$ dependence of $\G $ and $ \D$ at this stage in calculating $F$. Integrating two more times, we get,
\be
 \frac{\partial F}{\partial P} = rP + \frac{1}{3!}(\G+\D)P^3.
\label{1-diff}
\ee
To obtain a form of $F$ suitable to describe the first order transition we need to retain the $P$ dependence of $\G$ and $\D$ at this stage and thus,
\bea
F&=& \int_0^P (rP' + \frac{1}{3!}(\G+\D)P'^3)dP' \nonumber\\
&=& \frac{1}{2}r P^2 +\frac{(\G(P)+\D(P))}{4!}P^4\nonumber\\
&&-\frac{1}{4!}\int_0^P P'^4\frac{d}{dP'}(\G(P')+\D(P'))dP'.
\label{f-int}
\eea
To evaluate the above integral, we make the following substitution(using eqn. (\ref{parquet}))
\be 
P^2 =\frac{\Lambda e^{-x}-r}{\frac{\G}{2}+\frac{\D}{6}}.
\ee
Contribution from the integral part in the previous equation is given by
\bea
&-&\frac{1}{4!}\int_{\log(\Lambda/r)}^{\log\frac{\Lambda}{r+(\frac{\G}{2}+\frac{\D}{6})P^2}} \lb\frac{\Lambda e^{-x}-r}{(\frac{\G}{2}+\frac{\D}{6})}\rb^2\times \frac{d}{dx}(\G(x)+\D(x))dx\nonumber\\
&=& \frac{\K}{4}\int_{\log(\Lambda/r)}^{\log\frac{\Lambda}{r+(\frac{\G}{2}+\frac{\D}{6})P^2}}
(\Lambda^2e^{-2x}-2r\Lambda e^{-x}+r^2)dx\nonumber\\
&=&  \frac{\K}{4}\lb\frac{r}{2}(\G+\D)P^2-\frac{1}{2}\lb\frac{\G}{2}+\frac{\D}{6}\rb^2\rb P^4\nonumber\\
&&+r^2\log\frac{r}{r+(\frac{\G}{2}+\frac{\D}{6})P^2}).
\eea
Thus we get the following expression for the free energy at zero temperature
\bea
F &=& \frac{\K}{4}r^2\log\frac{6r}{6r+(3\G+\D)P^2}+\frac{1}{2}rP^2\lb 1+\frac{\K}{2}\lb\frac{\G}{2}+\frac{\D}{6}\rb\rb\nonumber \\
&&+ P^4\lb\frac{\G+\D}{4!}-\frac{\K}{8}\lb\frac{\G}{2}+\frac{\D}{6}\rb^2\rb.
\label{FE}
\eea
where $\G$ and $\D$ are $P$-dependent and are to be determined from the set of equations (\ref{self-eqn1},\ref{self-eqn2}). We notice that only the following 
combinations of $\G$ and $\D$ appear in all the calculations,
\bea 
\gamma_1 &=& 3\G+\D, 
\gamma_2  = \G+\D. 
\eea
Here the bare value of $\gamma_2$ ($\gamma^0_2$)is the coefficient for the quartic term in mean field approximation. On the other hand non-zero polarization enters into the fluctuation propagator with a coupling constant $\gamma_1$. In case of $\gamma^0_2<0$, a mean field picture requires an additional $|P|^6$ term with positive coefficient for the stability of the system. However a fluctuation corrected scenario can ensure stability without such a term, provided $\gamma_1>0$. With the above definitions, the set of equations (\ref{gamma-delta}) becomes a single self-consistent equation for $\gamma_1$ 
 \bea
\gamma_1 &=&  \frac{3\G^0}{1+ \frac{3}{2}\K \G^0 x}\lb 1+\frac{\D^0}{\D^0+(3\G^0-\D^0)(1+ \frac{3}{2}\K \G^0 x)}\rb.
\eea
Here $x$ contains $\gamma_1$ only. Solving the above equation, $\gamma_2$ can be found from (eqn. \ref{self-eqn2})
\be
 \frac{d\gamma_2}{dx} = -\frac{\K}{6}\gamma_1^2.
\label{diff-g2}
\ee
Below $\gamma_2=0$, phase transition in this system will be first order. We are interested in the phase transition near $\gamma_2= 0$, where fluctuation effects in four point vertices are important. If we limit ourselves to the region $|\D|< 3\G$, the leading order behavior of $\gamma_1$ is same as that of $\G$ and is given by
\bea
\gamma_1 &\approx&  \frac{3\G^0}{1+ \frac{3}{2}\K \G^0 x}=\frac{3\G^0}{1+ \frac{3}{2}\K \G^0 \log\frac{6\Lambda}{\sqrt{6r+\gamma_1P^2}}}.
\eea 
Assuming the bare value $\G^0\sim \mathcal{O}(10)$, so that $\frac{\G^0}{4!}<1$ which validates perturbation theory, 
gives $6\G^0\sim \mathcal{O}(10^{2})$. If we define a parameter $a= \frac{3}{4}\K \G^0$, then within the validity regime of perturbation theory $a\sim\mathcal{O}(10^{-1})$. 
Thus for an wide range of $\gamma_1$ e. g. $\gamma_1\sim \mathcal{O}(1)-\mathcal{O}(10^2)$, we find $|a\log \gamma_1|<|\frac{3\G^0}{\gamma_1}|$. For small $r$, $\gamma_1$ and $\gamma_2$ can be estimated as
\bea 
\gamma_1\sim\frac{3\G^0}{1-a\log(\gamma_1P^2)}
\approx\frac{3\G^0}{1-2a\log P}.
\label{g1-T0}
\eea
This is essentially a non-self-consistent solution for $\gamma_1$. Variation of $\gamma_1$ at zero temperature is shown in figure \ref{zero}. From the figure it is visible that, at $P=0$ because of the quantum critical fluctuations there is a strong reduction of $\gamma_1$ from its bare value and a non-zero $P$ restores it to its bare value. A non-zero polarization has similar effects on the $\gamma_2$ which leads to a first order transition.

From the equation (\ref{diff-g2}), we get,
\bea
\gamma_2 &=& \gamma_2^0+(-9(\G^0)^2+\frac{\gamma_1}{3})\label{g2}\\
&=& \gamma_2^0-\frac{9(\G^0)^2\K}{6a}\lb1-\frac{1}{1-2a\log P}\rb
\eea
where $\gamma_2^0$ is the bare value of $\gamma_2$. Since the corrections due to self-consistency $\sim \log \gamma_1$, the above estimate breaks down near $P\sim \exp(1/2a)$. Except in that regime, the non-self-consistent result is expected to give good result.
\figyz

{\bf Free energy at zero temperature:} Using the asymptotic behavior of the four point vertices (eqn. (\ref{g1-T0})) and 
using equation (\ref{FE}) defining $\tilde{\gamma}_2^0$ as $\gamma_2^0-\frac{9(\G^0)^2\K}{6a}$, we can write the following asymptotic expression for the free energy at zero temperature
\bea
F &=& \frac{\K}{4}r^2\log\frac{6r}{6r+P^2\frac{3\G^0}{1-2a\log P}}+\frac{1}{2}rP^2\lb1+\frac{\K}{12}\frac{6\G^0}{1-2a\log P}\rb\nonumber \\
&&+ \frac{ P^4}{4!}\lb\lb\tilde{\gamma}_2^0+\frac{\K}{6a}\lb\frac{9(\G^0)^2}{1-2a\log P}\rb \rb-\frac{4!\K}{36\times 8}\lb\frac{3\G^0}{1-2a\log P}\rb^2\rb.
\label{FE-T0} 
\eea
Here the first term $\sim\mathcal{O}(r^2)$, hence negligible compared to the other terms in the vicinity of a quantum critical point($r\rightarrow 0$). The second term is a standard quadratic term with fluctuation corrections. The third term, describes the appropriate physics of the problem. The coefficient of the quartic term 
contains three terms. First one is a constant and can take either positive or negative but small values. Now there is a
competition between the second and the third term. The second term tries to make the free energy positive while the third term tries 
to make it negative. However unless $\gamma_2^0$ is a sufficiently large negative number, coefficient of the quartic term is positive in the parameter regime of our interest.

To have a phase transition, the following equation must be satisfied,
\bea
\frac{\partial F}{\partial P}&=&rP+\frac{1}{3!}(\G+\D)P^3=0
\eea
For positive $r\approx 0$, the above equation can be satisfied only if $\G+\D$ is negative. Corresponding value of the $P$ 
is given as,
\be
 P_0=\exp\lb \frac{1}{2a}+\frac{9\K(\G^0)^2}{12a\tilde{\gamma}_2^0} \rb =\exp\frac{1}{2a}\lb1+\mu\rb
\label{P}
\ee
with $\mu=\frac{9\K(\G^0)^2}{6\gamma_2^0}$.
Since $|\mu|>>1$, the parameter $\tilde{\gamma}_2^0>0$ corresponds to the very large values of $P_0$. Since the scheme presented here, is valid for small $P$, we exclude this possibility in this discussion. On the other hand, $\tilde{\gamma}_2^0<0$ corresponds to the finite value of $P_0$ and there is a possibility of a first 
order transition at $r=r_0>0$. As $r\sim P^2$, $r_0\sim \exp\frac{1}{a}\lb1+\mu\rb$.  This is in sharp contrast to the mean-field prediction. In the later case a discontinuity in the order parameter is predicted to be $\sim \gamma_2^0=|u+v|$, while such discontinuity has a non-analytic dependence on $\gamma_2^0$ in a fluctuation induced and order parameter limited transition in our theory.
If we consider the Gaussian thermal fluctuations near the instability point and neglect the temperature dependencies of the vertices, then the thermal corrections to 
$r$ and $P^2$  $\sim T^2$ and $P_0^2$ should vanish above a temperature $T\sim \exp\frac{1}{2a}\lb1+\mu\rb$. 
However the fluctuation cut-off for the renormalized vertices and hence the form of fluctuation re-normalized free energy should 
get changed at finite temperature. In the next section we will discuss the finite temperature case in detail. 
\section{Fluctuation corrections to the free energy at finite temperature}
\figGap
 Here we consider the system to be at a low but non-zero temperature as well as near a mean field quantum critical
point with a zero temperature negative gap( i. e. $r=-r_0, \, r_0\ge 0$). Near $r_0=0$, fluctuation corrections at finite temperature to it leads to
\bea
r(T)=-r_0+K_3 \lb \frac{u}{2}+\frac{v}{6}\rb T^2.
\eea
Here the thermal fluctuations are considered up-to the Gaussian level as done in our earlier works\cite{we,mag}. Above expression for $r(T)$ without any correction 
to four point vertices predicts a second order transition with a transition temperature $T_c\sim \sqrt{r_0}$ for $u+v>0$. Again we
will look at the correction to the four point vertices in the limit $|u+v|\rightarrow 0$ but at non-zero temperature.
Using the same procedure as used for the zero temperature case, we now deduce the parquet equations for the four point vertices and hence the free energy at finite temperature. 
At non-zero temperature, fluctuation corrections to the four point vertices takes the form
\bea
\delta u &\sim&   -u^2 T \int_0^\Lambda d^{3}q G^2(q,~\omega_n=0)\nonumber\\
&\sim&  -u^2 K_3\frac{T}{(r(T)+ (u/2+v/6)P^2)^{\frac{1}{2}}},
\eea
in three dimension. Near the critical point, $r(T) \sim r + K_3(u/2+v/6)T^2$. Thus at finite temperature, we can define finite temperature fluctuation cut-off as
\be
x_T =  \frac{T}{(r+ (u/2+v/6)(P^2+K_3T^2))^{\frac{1}{2}}}.
\ee
In this case, fluctuation corrections to the free energy in terms of $\gamma_1$ and $\gamma_2$ (equation \ref{f-int}) becomes,
\bea
I_{fluc}&=&-\frac{1}{4!}\int_{ x_T(0)}^{ x_T(P)} \lb\frac{T^2/x^2-r}{{\gamma_1}/{6}} -K_3T^2\rb^2 \frac{d\gamma_2(x)}{dx}dx\nonumber \\
&=& \frac{1}{4}[ T[(r+ \frac{\gamma_1}{6}K_3T^2)^{\frac{3}{2}}-(r+ \frac{\gamma_1}{6}(P^2+K_3T^2))^{\frac{3}{2}}]\nonumber\\
&&+r^2 [(r+ \frac{\gamma_1}{6}K_3T^2)^{-\frac{1}{2}}
-(r+ \frac{\gamma_1}{6}(P^2+K_3T^2))^{-\frac{1}{2}}]\nonumber \\
&&+ T^4\gamma_2^2 [(r+ \frac{\gamma_1}{6}K_3T^2)^{-\frac{1}{2}}-(r+ \frac{\gamma_1}{6}(P^2+K_3T^2))^{-\frac{1}{2}}]\nonumber\\
&&- Tr[(r+ \frac{\gamma_1}{6}K_3T^2)^{\frac{1}{2}}-(r+ \frac{\gamma_1}{6}(P^2+K_3T^2))^{\frac{1}{2}}]\nonumber \\
&&+ 2rT^2\gamma_2[(r+ \frac{\gamma_1}{6}K_3T^2)^{-\frac{1}{2}}-(r+ \frac{\gamma_1}{6}(P^2+K_3T^2))^{-\frac{1}{2}}]\nonumber \\
&&+ T^4\gamma_2[(r+ \frac{\gamma_1}{6}K_3T^2)^{-\frac{1}{2}}-(r+ \frac{\gamma_1}{6}(P^2+K_3T^2))^{-\frac{1}{2}}]].
\eea
Here $x_T(P) = \frac{T}{(r+ \frac{\gamma_1}{6}(P^2+K_3T^2))^{\frac{1}{2}}}$ is used as polarization dependent fluctuation cut-off at non-zero temperature.  In performing the above integral, $x_T$-dependence of $\G$ and $\D$ are neglected as that would lead to sub leading corrections. For the systems near $r\rightarrow 0$ limit, 
retaining only the terms lowest order in $T$ and $r$, the free energy can be truncated as
\bea
F&=& \frac{1}{2}rP^2+ P^4\lb\frac{\G+\D}{4!}\rb+\frac{T}{4}[\lb r+ K_3(\frac{\G}{2}+\frac{\D}{6})T^2\rb^{\frac{3}{2}}\nonumber \\
&&- \lb r+ (\frac{\G}{2}+\frac{\D}{6})(P^2+K_3T^2)\rb^{\frac{3}{2}}].
\eea
For small $r$ and $T$, the third term is of the $\mathcal{O}(T^{4})$, hence is negligible. Thus the free energy in the 
leading order can be further truncated as
\bea
F=\frac{r}{2}P^2+ P^4\lb \frac{\G+\D}{4!}\rb -\frac{T}{4}\lb \frac{\G}{2}+\frac{\D}{6}\rb^{\frac{3}{2}}P^3.
\label{fe}
\eea
In the above equation $P\equiv|\vec{P}|$ and the cubic term which is a result of  small $P$ expansion, does not violate the symmetry of the problem. A finite temperature version of the equation (\ref{g1-T0}), i.e. the asymptotic form of the self consistent equation for $\gamma_1$, one of the important combinations of the four point vertices reads,
\be
 \gamma_1=\frac{6\G^0}{1-\frac{6aT}{(\gamma_1(P^2+K_3T^2))^{1/2}}}.
\ee
Non-zero solution of the above equation tells
\bea
\gamma_1 &=& \frac{18a^2T^2+K_3\G^0T^2+\G^0P^2}{P^2+K_3T^2}\pm \frac{6aT\sqrt{9  a^2T^2+ \G^0P^2 + \G^0K_3T^2}}{P^2+K_3T^2}.
\eea
In deriving the equations at finite temperature we have limited ourselves in the low temperature region i.e.  $a^2T^4$ is neglected compared to $K_3T^2$. Moreover $\gamma_1$ should be strongly suppressed due to critical fluctuations at zero $P$ and should go towards its bare value with increasing $P$. Hence the part is remaining,
\be 
\gamma_1\approx \G^0-\frac{6a\sqrt{\G^0}T}{\sqrt{P^2+K_3T^2}}.
\label{yT}
\ee
Asymptotic evolution of $\gamma_1$ with $P$ for two different temperature is shown in figure \ref{finite}. In this figure we find that qualitative nature of the curve is similar to that of the zero temperature case except that at $P=0$, reduction of $\gamma_1$ at a finite temperature is lower than that of it at a zero temperature. Since at a finite temperature the quantum critical fluctuations are suppressed, this is an expected result. 

However from equation (\ref{g2}) we get
\be
\gamma_2= \gamma_2^0+\lb\frac{2\G^0}{3}+\frac{2a\sqrt{\G^0}T}{\sqrt{P^2+K_3T^2}}\rb.
\ee
\figT

{\bf Free energy at finite temperature:} Again using the asymptotic behavior of the four point vertices (\ref{yT}) and using 
equation (\ref{fe}), we can write the following asymptotic expression for the free energy at non zero temperature
\bea
F&=& \frac{r(T)}{2}P^2
+\frac{P^4}{4!}\lb\gamma_2^0+\lb\frac{2\G^0}{3}+\frac{2a\sqrt{\G^0}T}{\sqrt{P^2+K_3T^2}}\rb\rb\nonumber\\
&&-\frac{T}{4}P^3\lb\frac{\G^0}{6}-\frac{a\sqrt{\G^0}T}{\sqrt{P^2+K_3T^2}}\rb^{\frac{3}{2}}.
\label{FE-T} 
\eea
Gap renormalization up-to one loop(figure \ref{gap}) at finite temperature near $r=0$, tells,
\be
r(T)= -r_0+K_3\lb\frac{\G^0}{6}-\frac{a\sqrt{\G^0}T}{\sqrt{P^2+K_3T^2}} \rb T^2.
\ee
Neglecting the term of $\mathcal{O}(T^{-3})$ in the coefficient of $P^2$, the expression for free energy becomes
\bea
F&=& \frac{1}{2}\lb-r_0+\frac{K_3\G^0}{6}T^2\rb P^2\nonumber \\
&&+\frac{1}{4!}P^4\lb \gamma_2^0+\lb\frac{2\G^0}{3}+\frac{2a\sqrt{\G^0}T}{\sqrt{P^2+K_3T^2}}\rb\rb\nonumber\\
&&-\G^0\frac{T}{4}P^3\lb\frac{1}{6}-\frac{3aT}{2\sqrt{\G^0(P^2+K_3T^2)}}\rb. 
\eea 
In the limit $P^2>>K_3T^2$, it takes the form
\bea
F&=& \frac{1}{2}\lb-r_0+\frac{K_3\G^0}{6}T^2+\frac{3aT^2\sqrt{\G^0}}{2}\rb P^2\nonumber\\
&&-\frac{T}{24}P^3\lb\G^0-\frac{a}{2}\sqrt{\G^0}\rb+\lb\frac{\frac{2\G^0}{3}+\gamma_2^0}{4!}\rb P^4.
\label{FE-TR}
\eea
This assumption does not contradict $K_3 T^2$ contribution to the gap renormalization. Such contribution appeared with the assumption that the lowest non-zero Matsubara frequency, i.e. $2\pi T> \sqrt{\frac{\gamma_1}{6}} P$. This assumption holds good even if $P^2>>K_3T^2$. Above equation tells that, the solution of the equation
${\partial F}/{\partial P} =0$ will result a nonzero value of $P_0\sim T_0$, even if $\gamma_2^0>0$. Here $T_0$ 
is the second order transition temperature for the mean field theory. Since near quantum critical point, $T_0\sim \sqrt{r_0}$, $P_0$ is 
also $\sim \sqrt{r_0}$. If we compare this results with that of the zero temperature case, we see that  in the finite temperature case, the discontinuity in the order parameter near the transition point $\sim \sqrt{r_0}$ while it is independent of $r_0$ in the zero temperature case. Since $r_0$ is the smallest scale(near mean field quantum critical point, $r_0\rightarrow0$) in this system, first order transition as a result of the order parameter fluctuations at finite temperature near the mean field quantum critical point is weaker than that of the zero temperature case.  This result is consistent with some experimental fact as reported in reference\cite{Taniguchi}.
\section{Discussions} Occurrence of first order transition due to  coupling between the order parameter fluctuations and soft modes is a well studied problem in many classical systems\cite{Larkin,pokrovsky,rudnick,Amit} and recently studied in context of quantum phase transitions in electronic systems\cite{Belitz}. However fluctuation effects in a weak first order transition, its dependence on effective dimension is not well studied in the literature of quantum phase transitions. We attempt the same here and make some connections with relevant experimental findings. To capture the basic physics near a weak first order transitions, one needs to study the effects of fluctuations on the proper four point vertices. Bare perturbation calculations in this system show vertex corrections to be logarithmically singular at zero temperature and with zero polarization in three dimensions due to quantum critical fluctuations. To include the effects of the singular contributions, we use the lowest order renormalization 
group equations to derive a set of recursion relations for four point vertices. Moreover to stabilize the system a non-zero polarization is assumed. Here the re-normalized vertices crucially depend on the non-zero polarization. Using these relations, the 
expressions for the free energy both at zero temperature and at finite temperature are derived. 
 The relevant quantities like transition temperature and the discontinuity in the order parameter at the transition point turn out to be small but finite.  We mainly concentrate on the phenomena near a \emph{ quantum critical point} predicted by a mean field theory. We found stronger possibility of first order transitions at $T=0$ than at any finite temperature transition near a mean field quantum critical point. This can certainly be related to the effective dimension of the dielectric fluctuations is four, i.e. marginal here. 
 From our analysis it is found that a finite polarization is required to  suppress the effects of the dielectric fluctuations at zero temperature where the system is at its marginal dimension. Since the effects of long wavelength dielectric fluctuations at marginal dimension get reduced due to finite temperature, a lower value is sufficient to stabilize the system. Thus in our scenario, the phenomena observed in the experiment\cite{Taniguchi} is  a consequence of the quantum critical fluctuations at its marginal dimension.  Apart from that, unlike the standard renormalization group approach, 
which considers first order transition as just the inability of the system to reach an unstable fixed point, the present approach makes qualitative predictions about the magnitude of the discontinuity in order-parameter near the transition point. The formalism presented here is quite general and can be used in case of first order transitions where order parameter fluctuations are important and correlation length is large though not divergent. 
We have considered the constant pressure case only, where strain fluctuations are completely integrated out. One can also consider a situation where system volume is constant\cite{Bergman} and a proper quantum generalization of such case should certainly be addressed.

\end{document}